\documentclass[intlimits,twoside,a4paper]{article}

\usepackage{amsmath,amssymb}
\usepackage{graphicx}
\usepackage{wrapfig}

\usepackage[T2A]{fontenc}
\usepackage[cp1251]{inputenc}
\providecommand{\abs}[1]{\lvert#1\rvert}
\DeclareMathOperator{\Sp}{Sp}

\usepackage[eqsecnum]{cmpj}

\issue{2011}{14}{1}{13004}

\doinumber{10.5488/CMP.14.13004}

\title[Two-state Bose-Hubbard model in the HCB limit]%
{Two-state Bose-Hubbard model \\ in the hard-core  boson limit}
\author{I.V. Stasyuk, O.V. Velychko}
\address{Institute for Condensed Matter Physics
of the National Academy of Sciences of Ukraine, 1~Svientsitskii
Str., 79011 Lviv, Ukraine}
\begin{document}

\maketitle

\begin{abstract}

Phase transition into the phase with Bose-Einstein (BE) condensate
in the two-band Bose-Hubbard model with the particle hopping in
the excited band only is investigated. Instability connected with
such a transition (which appears at excitation energies
$\delta<\lvert t_0' \rvert$, where $\lvert t_0' \rvert$ is the
particle hopping parameter) is considered. The re-entrant
behaviour of spinodales is revealed in the hard-core boson limit
in the region of positive values of chemical potential. It is
found that the order of the phase transition undergoes a change in
this case and becomes the first one; the re-entrant transition
into the normal phase does not take place in reality. First order
phase transitions  also exist at negative values of $\delta$ (under
the condition $\delta>\delta_{\mathrm{crit}}\approx-0.12\lvert
t_0' \rvert$). At $\mu<0$ the phase transition mostly remains to
be of the second order.
The behaviour of the BE-condensate order parameter is analyzed,
the $(\Theta,\mu)$ and $(\lvert t_0' \rvert,\mu)$ phase diagrams
are built and localizations of tricritical points are established.
The conditions are found at which the separation on the normal
phase and the phase with the BE condensate takes place.
\keywords Bose-Hubbard model, hard-core bosons, Bose-Einstein
condensation, excited band
\pacs 03.75.Hh, 03.75.Lm, 64.70.Tg, 71.35.Lk, 37.10.Jk, 67.85.-d
%
%
%
%
%
%
%
%
%
%
%
\end{abstract}

\section{Introduction}

During the recent years Bose-Hubbard model (BHM) is proved to be a
valuable tool in the theory of systems of strongly correlated
particles. The model achieves a wide recognition due to a
successful description of thermodynamics and dynamics of ultracold
Bose atoms in optical lattices where a phase transition to the
phase with the Bose-Einstein (BE) condensate (so-called Mott
insulator (MI) -- superfluid state (SF) transition) occurs at very
low temperatures. Experimental evidence of BE condensation in
optical lattices was found for the first time in
works~\cite{wrk01,wrk02} while theoretical predictions of such an
effect were given earlier~\cite{wrk03}. Starting from the 90-ies
of the past century, a series of papers were devoted to the theory
of this phenomenon. Among the first key articles on the subject
one should mention the work~\cite{wrk04} where BHM was studied in
the mean field approximation. The calculated therein phase
diagrams demonstrate that in the simplest case (i.e., hopping of
Bose particles in the presence of a single-site Hubbard repulsion)
the MI-SF transition is of the second order. Moreover, it is
supposed that particles reside in the ground state of local
potential wells in the lattice. Forthcoming theoretical
investigations in this field were performed with the use of
various techniques, e.g., the random phase approximation (RPA) in
the Green function method~\cite{wrk05,wrk06}, a strong-coupling
perturbation theory~\cite{wrk07,wrk08}, the dynamical mean field
theory (Bose-DMFT)~\cite{wrk11,wrk12} as well as quantum
Monte-Carlo calculations~\cite{wrk09,wrk10} and other numerical
methods.

The Bose-Hubbard model is also intensively used for a theoretical
description of a wide range of phenomena: quantum delocalization
of hydrogen atoms adsorbed on the surface of transition
metals~\cite{wrk13,wrk14}, quantum diffusion of light particles on
the surface or in the bulk~\cite{wrk15,wrk16}, thermodynamics of
the impurity ion intercalation into
semiconductors~\cite{wrk17,wrk18}.

In the last mentioned applications, there is usually a restriction
on the position occupation number ($n_i\leqslant1$), which  corresponds to the limit of an infinite Hubbard
repulsion for the
considered model. Such a model of hard-core ions (where particles are
described by the Pauli statistics) is  also known as the
fundamental one for the investigation of a wide range of problems,
e.g.,\ superconductivity due to a local electron pairing
\cite{wrk19} or ionic hopping in ionic (superionic) conductors
\cite{wrk20,wrk21}.

The study of a quantum delocalization or diffusion reveals an
important role of excited vibrational states of particles (ions)
in localized (interstitial) positions with a much higher
probability of ion hopping between them~\cite{wrk15,wrk22,wrk23}.
A similar issue of a possible BE condensation in the excited bands
in optical lattices is also considered but the condition of their
sufficient occupation due to the optical pumping (see, e.g.\
\cite{wrk24}) is imposed. An orbital degeneration of the excited
$p$-state is accompanied by anisotropy of hopping parameters and
causes the appearance of variously polarized bands in the
one-particle spectrum. Such bands correspond by convention to
different sorts (so-called ``flavours'') of bosons and their
number correlates with the lattice dimensionality. In the
framework of the necessary generalization of the Bose-Hubbard
model, a possibility of the MI-SF transition to the phase with BE
condensate in the pumping-induced quasi-equilibrium long-living
state of the system has been established~\cite{wrk25}.

In the equilibrium case, the issue of BE condensation involving
the excited states in the framework of ordinary Bose-Hubbard model
was not considered in practice. The exception is the system of
spin-1 bosons~\cite{wrk26,wrk27} where a hyperfine splitting gives
rise to multiplets of local states resulting in closely-spaced
excited levels. As demonstrated in~\cite{wrk28,wrk29}, the MI-SF
phase transition could be of the first order when a single-site
spin interaction is of the ``antiferromagnetic'' type. A similar
change of the phase transition order also takes place  for
multicomponent Bose systems in the optical lattices~\cite{wrk30}.

In the present work we consider an equilibrium thermodynamics of
the Bose-Hubbard model taking into account only one nondegenerated
excited state on the lattice site besides the ground one. On the
one hand, such a model corresponds to 1D or strongly anisotropic
(quasi-1D) optical lattice, and on the other hand, it is close to a
situation that is characteristic of a system of light particles
adsorbed on the metal surface. For example, the excited states of
hydrogen atoms on the Ni(111) surface are sufficiently distant
\cite{wrk22} so only the lowest one could be taken into account.
We shall investigate a condition of instability of a normal state
of the Bose system with respect to BE condensation considering a
criterion of divergence of the susceptibility
($\chi\sim\langle\langle c_l | c_p^+
\rangle\rangle_{\omega}|_{q=0,\omega=0}$)
characterizing the system response with respect to the field
related to a spontaneous creation or annihilation of particles. We
shall also study the behaviour of the order parameter $\langle c_0
\rangle$ ($\langle c_0^+ \rangle$) as well as the grand canonical
potential in the region of the MI-SF transition and shall build
relevant phase diagrams. Special attention will be paid to a
change of the phase transition order and localization of
tricritical points at different values of excitation energy,
particle hopping parameter and temperature.

We shall limit ourselves to the hard-core boson (HCB) limit where
a limitation on occupation numbers is present: no more than one
particle per site regardless of the state (excited or ground) occupied by it. Thus, the single-site problem is a three-level one
(contrary to the two-level ordinary HCB case). For this reason, it
is convenient to use the formalism of Hubbard operators~\cite{wrk31} (standard basis operators~\cite{wrk32}).

\section{Two-state Bose-Hubbard model in RPA:
normal phase}

The Bose-Hubbard model is used for description of the
system of Bose particles which are located in a periodic field and
can reside in lattice sites. Taking into account only the ground
and the first excited vibrational levels in the potential well on
the site, one can express the model Hamiltonian as:
\begin{align}
    \hat{H} &= (\varepsilon-\mu)\sum_i b_i^+ b_i
        + (\varepsilon'-\mu)\sum_i c_i^+ c_i
        + \frac{U_b}{2}\sum_i n_i^b(n_i^b-1)
        + \frac{U_c}{2}\sum_i n_i^c(n_i^c-1)
        + U_{bc}\sum_i n_i^b n_i^c
    \notag\\
    &\quad + \sum_{ij} t_{ij}^b b_i^+ b_j
        + \sum_{ij} t_{ij}^c c_i^+ c_j
        + \sum_{ij} t_{ij}^{bc} (b_i^+ c_j + c_i^+ b_j),
    \label{eq2-01}
\end{align}
where $b_i$ and $b_i^+$ ($c_i$ and $c_i^+$) are Bose operators of
annihilation and creation of particles in the ground (excited)
state, $\varepsilon$ and $\varepsilon'$ are respective energies of
state and $\mu$ is the chemical potential of particles. Such a
Hamiltonian includes the single-site Hubbard repulsions with
energies $U_b$, $U_c$ and $U_{bc}$ as well as the particle hopping
between ground ($t^b$), excited ($t^c$) and different ($t^{bc}$)
states. Hereinafter we assume $U_b=U_c=U_{bc}$ for simplicity.

Let us define a single-site basis
$|n_i^b,n_i^c\rangle\equiv|i;n_i^b,n_i^c\rangle$
(which is formed by particle occupation numbers in the ground and in
the excited states, i.e., eigenvalues of operators
$n_i^b=b_i^+ b_i$ and $n_i^c=c_i^+ c_i$)
as well as introduce Hubbard operators (standard basis operators)
\begin{equation}
    X_i^{n,m;n',m'} \equiv |i;n,m\rangle\langle i;n',m'|.
    \label{eq2-02}
\end{equation}
Annihilation and creation Bose operators may be written as
\begin{align}
    b_i &= \sum_n\sum_m \sqrt{n+1}\, X_i^{n,m;n+1,m},
    &
    b_i^+ &= \sum_n\sum_m \sqrt{n+1}\, X_i^{n+1,m;n,m};
    \notag\\
    c_i &= \sum_n\sum_m \sqrt{m+1}\, X_i^{n,m;n,m+1},
    &
    c_i^+ &= \sum_n\sum_m \sqrt{m+1}\, X_i^{n,m+1;n,m}.
    \label{eq2-03}
\end{align}
Corresponding occupation numbers look as follows
\begin{align}
    n_i^b &= \sum_n\sum_m n X_i^{n,m;n,m},
    &
    n_i^c &= \sum_n\sum_m m X_i^{n,m;n,m},
    \label{eq2-04}
\end{align}
where summation indices $n,m=0,\dotsc,\infty$ in both
\eqref{eq2-03} and \eqref{eq2-04} formulae.

In the $X$-operator representation, the single-site part of
Hamiltonian \eqref{eq2-01} can be written as
\begin{align}
    \hat{H}_0 &= \sum_i\sum_n\sum_m \lambda_{nm} X_i^{n,m;n,m},
    \label{eq2-05}\\
    \intertext{where}
    \lambda_{nm} &= n(\varepsilon-\mu)+m(\varepsilon'-\mu)
        +\frac{U}{2}(n+m)(n+m-1).
    \label{eq2-06}
\end{align}
Terms describing an inter-site transfer in Hamiltonian
\eqref{eq2-01} are transformed in a similar way.

Our primary goal is to calculate the two-time temperature boson
Green's functions
$\langle\langle b | b^+ \rangle\rangle$
and
$\langle\langle c | c^+ \rangle\rangle$,
which describe an excitation spectrum and make it possible to
investigate the conditions of the system's instability with
respect to the spontaneous symmetry breaking and the appearance of
a BE condensate. As follows from definitions \eqref{eq2-03}
\begin{align}
    \langle\langle b_l | b_p^+ \rangle\rangle_{\omega}
    &=
    \sum_{nm}\sum_{rs} \sqrt{n+1}\,\sqrt{r+1}\,
    \langle\langle
        X_l^{n,m;n+1,m} | X_p^{r+1,s;r,s}
    \rangle\rangle_{\omega}\,,
    \notag\\
    \langle\langle c_l | c_p^+ \rangle\rangle_{\omega}
    &=
    \sum_{nm}\sum_{rs} \sqrt{m+1}\,\sqrt{s+1}\,
    \langle\langle
        X_l^{n,m;n,m+1} | X_p^{r,s+1;r,s}
    \rangle\rangle_{\omega}\,.
    \label{eq2-07}
\end{align}

We will use the equation-of-motion method for the evaluation of
$X$-operator Green's functions. For the first one, from relations
\eqref{eq2-07} one could write
\begin{align}
    \hbar\omega
    \langle\langle
        X_l^{n,m;n+1,m} | X_p^{r+1,s;r,s}
    \rangle\rangle_{\omega}
    &=
    \frac{\hbar}{2\pi}
    \langle
        X_l^{n,m;n,m} - X_l^{n+1,m;n+1,m}
    \rangle
    \delta_{lp}\delta_{nr}\delta_{ms}
    \notag\\
    &\quad
    +
    \langle\langle
        [X_l^{n,m;n+1,m},\hat{H}] | X_p^{r+1,s;r,s}
    \rangle\rangle_{\omega}\,.
    \label{eq2-08}
\end{align}
Let us write the commutators
\begin{align}
    [X_l^{n,m;n+1,m},\hat{H}_0] &= (\lambda_{n+1,m}-\lambda_{n,m})
        X_l^{n,m;n+1,m},
    \label{eq2-09}\\
    \refstepcounter{equation}
    [X_l^{n,m;n+1,m},b_i^+] &= \delta_{li}
        \sqrt{n+1} \left(X_l^{n,m;n,m}-X_l^{n+1,m;n+1,m}\right),
    \tag{\theequation{}a}\label{eq2-10a}\\
    [X_l^{n,m;n+1,m},b_i] &= \delta_{li}
        \left(\sqrt{n+2}\, X_l^{n,m;n+2,m}
        - \sqrt{n}\, X_l^{n-1,m;n+1,m}\right),
    \tag{\theequation{}b}\label{eq2-10b}\\
    [X_l^{n,m;n+1,m},c_i^+] &= \delta_{li}
        \left(\sqrt{m}\, X_l^{n,m;n+1,m-1}
        - \sqrt{m+1}\, X_l^{n,m+1;n+1,m}\right),
    \tag{\theequation{}c}\label{eq2-10c}\\
    [X_l^{n,m;n+1,m},c_i] &= \delta_{li}
        \left(\sqrt{m+1}\, X_l^{n,m;n+1,m+1}
        - \sqrt{m}\, X_l^{n,m-1;n+1,m}\right).
    \tag{\theequation{}d}\label{eq2-10d}
\end{align}
The latter are originated from the commutation of an initial
$X$-operator with the inter-site transfer terms of the Hamiltonian,
thus producing the higher-order Green's functions
\begin{equation}
    \langle\langle
        X_l^{\dotsm} b_j | X_p^{r+1,s;r,s}
    \rangle\rangle_{\omega}
    \, ,\quad
    \langle\langle
        X_l^{\dotsm} b_j^+ | X_p^{r+1,s;r,s}
    \rangle\rangle_{\omega}
    \, ,\dotsc,
    \label{eq2-11}
\end{equation}
where $X_l^{\dotsm}$ stands for operators on the right-hand side
of expressions \eqref{eq2-10a}--\eqref{eq2-10d}.

Decoupling of functions \eqref{eq2-11} in the random phase
approximation (RPA) is performed in the following way:
\begin{equation}
    \langle\langle
        X_l^{\dotsm} b_j | X_p^{r+1,s;r,s}
    \rangle\rangle_{\omega}
    \approx
    \langle
        X_l^{\dotsm}
    \rangle
    \langle\langle
        b_j | X_p^{r+1,s;r,s}
    \rangle\rangle_{\omega}
    +
    \langle
        b_j
    \rangle
    \langle\langle
        X_l^{\dotsm} | X_p^{r+1,s;r,s}
    \rangle\rangle_{\omega}
    \, .
    \label{eq2-12}
\end{equation}
In the case of the normal phase (which will be studied herein)
$\langle b_j \rangle=\langle b_j^+ \rangle=0$. Thus, retaining
only the averages $\langle X_l^{\dotsm}\rangle$ of diagonal
$X$-operators we have
\begin{equation}
    [X_l^{n,m;n+1,m},\hat{H}]
    \approx
    \Delta_{nm} X_l^{n,m;n+1,m}
    +
    \sqrt{n+1}\, Q_{nm} \sum_j t_{lj} b_j
    +
    \sqrt{n+1}\, Q_{nm} \sum_j t_{lj}'' c_j
    \label{eq2-13}
\end{equation}
and equation \eqref{eq2-08} can be rewritten as
\begin{multline}
    \langle\langle
        X_l^{n,m;n+1,m} | X_p^{r+1,s;r,s}
    \rangle\rangle_{\omega}
    =
    \frac{\hbar}{2\pi} \delta_{lp}\delta_{nr}\delta_{ms}
    \frac{Q_{nm}}{\hbar\omega-\Delta_{nm}}
    \\
    +
    \frac{\sqrt{n+1}\, Q_{nm}}{\hbar\omega-\Delta_{nm}}
    \sum_j t_{lj}
    \langle\langle
        b_j | X_p^{r+1,s;r,s}
    \rangle\rangle_{\omega}
    +
    \frac{\sqrt{n+1}\, Q_{nm}}{\hbar\omega-\Delta_{nm}}
    \sum_j t_{lj}''
    \langle\langle
        c_j | X_p^{r+1,s;r,s}
    \rangle\rangle_{\omega}
    \,
    .
    \label{eq2-14}
\end{multline}
The following notations are introduced
\begin{equation}
    Q_{nm} = \langle X_l^{n,m;n,m} - X_l^{n+1,m;n+1,m} \rangle,
    \qquad
    \Delta_{nm} = \lambda_{n+1,m} - \lambda_{n,m}\,,
    \label{eq2-15}
\end{equation}
for the occupation difference of adjacent levels and the related
transition energies when the number of Bose particles in the
ground state (with the energy $\varepsilon$) on the site increases
by one.

Proceeding from $X$-operators in equation \eqref{eq2-14} to the Bose
operators $b$ and $b^+$ according to definition \eqref{eq2-03} we
obtain
\begin{equation}
    \langle\langle
        b_l | b_p^+
    \rangle\rangle_{\omega}
    =
    \frac{\hbar}{2\pi} \delta_{lp} g_0(\omega)
    +
    g_0(\omega)
    \biggl(
    \sum_j t_{lj}
    \langle\langle
        b_j | b_p^+
    \rangle\rangle_{\omega}
    +
    \sum_j t_{lj}''
    \langle\langle
        c_j | b_p^+
    \rangle\rangle_{\omega}
    \biggr)
    ,
    \label{eq2-16}
\end{equation}
where the function
\begin{equation}
    g_0(\omega)
    =
    \sum_{nm}
    \frac{Q_{nm}}{\hbar\omega-\Delta_{nm}}
    (n+1)
    \label{eq2-17}
\end{equation}
has the meaning of the unperturbed Green's function for bosons
residing in the single-site ground state.

Equations of motion for ``mixed'' Green's functions
$\langle\langle c | b^+ \rangle\rangle$
are obtained in the way similar to the  above described scheme.
Using decoupling \eqref{eq2-12} one can write
\begin{equation}
    [X_l^{n,m;n,m+1},\hat{H}]
    \approx
    \Delta_{nm}' X_l^{n,m;n,m+1}
    +
    \sqrt{m+1}\, Q_{nm}' \sum_j t_{lj}'' b_j
    +
    \sqrt{m+1}\, Q_{nm}' \sum_j t_{lj}' c_j
    \,,
    \label{eq2-18}
\end{equation}
which results in the equation
\begin{equation}
    \langle\langle
        c_l | b_p^+
    \rangle\rangle_{\omega}
    =
    g_0'(\omega)
    \biggl(
    \sum_j t_{lj}''
    \langle\langle
        b_j | b_p^+
    \rangle\rangle_{\omega}
    +
    \sum_j t_{lj}'
    \langle\langle
        c_j | b_p^+
    \rangle\rangle_{\omega}
    \biggr)
    .
    \label{eq2-19}
\end{equation}
Here, similarly to \eqref{eq2-15} and \eqref{eq2-17}
\begin{equation}
    Q_{nm}' = \langle X_l^{n,m;n,m} - X_l^{n,m+1;n,m+1} \rangle,
    \quad
    \Delta_{nm}' = \lambda_{n,m+1} - \lambda_{n,m}\,,
    \quad
    g_0'(\omega)
    =
    \sum_{nm}
    \frac{Q_{nm}'}{\hbar\omega-\Delta_{nm}'}
    (m+1),
    \label{eq2-20}
\end{equation}
and the function $g_0'(\omega)$ is the unperturbed Green's
function for bosons residing in the excited state.

By means of the Fourier transform
\begin{equation}
    \langle\langle
        b_l | b_p^+
    \rangle\rangle_{\omega}
    =
    \frac{1}{N}\sum_q \mathrm{e}^{\mathrm{i}\mathbf{q}(\mathbf{R}_l-\mathbf{R}_p)}
    \langle\langle
        b | b^+
    \rangle\rangle_{q,\omega}
    \,,
    \label{eq2-21}
\end{equation}
one can proceed to the momentum representation obtaining a system of
equations
\begin{align}
    \langle\langle
        b | b^+
    \rangle\rangle_{q,\omega}
    &=
    \frac{\hbar}{2\pi} g_0(\omega)
    +
    g_0(\omega)
    t_q
    \langle\langle
        b | b^+
    \rangle\rangle_{q,\omega}
    +
    g_0(\omega)
    t_q''
    \langle\langle
        c | b^+
    \rangle\rangle_{q,\omega}
    \,,
    \notag\\
    \langle\langle
        c | b^+
    \rangle\rangle_{q,\omega}
    &=
    g_0'(\omega)
    t_q''
    \langle\langle
        b | b^+
    \rangle\rangle_{q,\omega}
    +
    g_0'(\omega)
    t_q'
    \langle\langle
        c | b^+
    \rangle\rangle_{q,\omega}
    \,,
    \label{eq2-22}
\end{align}
where $t_q$\,, $t_q'$ and $t_q''$ stand for the Fourier transforms
of hopping parameters.

A pair of equations for other Green's functions are obtained in a
similar way
\begin{align}
    \langle\langle
        b | c^+
    \rangle\rangle_{q,\omega}
    &=
    g_0(\omega)
    t_q
    \langle\langle
        b | c^+
    \rangle\rangle_{q,\omega}
    +
    g_0(\omega)
    t_q''
    \langle\langle
        c | c^+
    \rangle\rangle_{q,\omega}
    \,,
    \notag\\
    \langle\langle
        c | c^+
    \rangle\rangle_{q,\omega}
    &=
    \frac{\hbar}{2\pi} g_0'(\omega)
    +
    g_0'(\omega)
    t_q''
    \langle\langle
        b | c^+
    \rangle\rangle_{q,\omega}
    +
    g_0'(\omega)
    t_q'
    \langle\langle
        c | c^+
    \rangle\rangle_{q,\omega}
    \,.
    \label{eq2-23}
\end{align}

Solutions of equations \eqref{eq2-22} and \eqref{eq2-23} are as follows:
\begin{align}
    \langle\langle
        b | b^+
    \rangle\rangle_{q,\omega}
    &=
    \frac{\hbar}{2\pi} \frac{1}{D_q(\omega)}
    g_0(\omega)(1-g_0'(\omega)t_q')
    ,
    \notag\\
    \langle\langle
        c | c^+
    \rangle\rangle_{q,\omega}
    &=
    \frac{\hbar}{2\pi} \frac{1}{D_q(\omega)}
    g_0'(\omega)(1-g_0'(\omega)t_q)
    ,
    \notag\\
    \langle\langle
        c | b^+
    \rangle\rangle_{q,\omega}
    &=
    \frac{\hbar}{2\pi} \frac{1}{D_q(\omega)}
    g_0(\omega)g_0'(\omega)t_q''
    =
    \langle\langle
        b | c^+
    \rangle\rangle_{q,\omega}
    \,,
    \label{eq2-24}
\end{align}
where
\begin{equation}
    D_q(\omega)
    =
    1-g_0(\omega)t_q-g_0'(\omega)t_q'+g_0(\omega)g_0'(\omega)
    \left[
        t_q t_q' - (t_q'')^2
    \right]
    .
    \label{eq2-25}
\end{equation}

The equation $D_q(\omega)=0$ gives the excitation spectrum which
is obtained here in the RPA. On the other hand, the divergence of
boson Green's functions \eqref{eq2-24} at the zero values of wave
vector and frequency is the criterion of instability with respect
to BE condensation~\cite{wrk05,wrk37}, thus giving the following
condition
\begin{equation}
    D_{q=0}(\omega=0)=0,
    \label{eq2-26}
\end{equation}
which can be rewritten in the explicit form
\begin{equation}
    1-g_0(\omega)t_q-g_0'(\omega)t_q'+g_0(\omega)g_0'(\omega)
    \left[
        t_q t_q' - (t_q'')^2
    \right]
    =0
    ,
    \label{eq2-27}
\end{equation}
where
\begin{equation}
    g_0(0)
    =
    -\sum_{nm}
    \frac{Q_{nm}}{(n+m)U-\mu}
    (n+1),
    \quad
    g_0'(0)
    =
    -\sum_{nm}
    \frac{Q_{nm}'}{(n+m)U+\delta-\mu}
    (m+1),
    \label{eq2-28}
\end{equation}
and $\delta=\varepsilon'-\varepsilon$ is the excitation energy.

We should point out that divergence of the
$\langle\langle b | b^+ \rangle\rangle_{0,0}$
function correlates with the appearance of the BE condensate in the
ground state while at the divergence of the
$\langle\langle c | c^+ \rangle\rangle_{0,0}$
function, BE condensation takes place in the excited state. In
general, both condensates appear simultaneously except the case
$t_q''=0$ (e.g.\ due to symmetry reasons) where these effects
become independent and only the one type of condensate arises in
the instability point.

Equation \eqref{eq2-27},  mutually relating the chemical potential,
hopping parameters and temperature, allows us to construct spinodal
surfaces (or lines) in the above mentioned coordinates and to find
the temperature of the phase transition to the phase with BE
condensate (so-called SF phase) where such a transition is of the
second order. Below, this problem (especially the issue of the
phase transition order) will be investigated  more in detail.

\section{NO phase instability in HCB limit}

Let us consider now a simple special case of the HCB limit when
occupation numbers in the $|n,m\rangle$ state are restricted by a
condition $n+m\leqslant1$. In the framework of the model, it
formally means $U\to\infty$.

In this case, the model becomes a three-level one with the local
energies
\begin{equation}
    \lambda_{00}=0,
    \qquad
    \lambda_{01}=\delta-\mu,
    \qquad
    \lambda_{10}=-\mu
    \label{eq3-01}
\end{equation}
and the following transition energies
\begin{equation}
    \Delta_{00}=-\mu,
    \qquad
    \Delta_{00}'=\delta-\mu.
    \label{eq3-02}
\end{equation}
Thus, equation \eqref{eq2-27} can be rewritten in the form
\begin{equation}
    1
    -\frac{Q_{00}}{\mu}t_0
    -\frac{Q_{00}'}{\mu-\delta}t_0'
    +\frac{Q_{00}Q_{00}'}{\mu(\mu-\delta)}
    \left[
        t_0 t_0' - (t_0'')^2
    \right]
    =0
    ,
    \label{eq3-03}
\end{equation}
where
\begin{equation}
    Q_{00}=
    \frac{1-\mathrm{e}^{\beta\mu}}
        {1+\mathrm{e}^{\beta\mu}+\mathrm{e}^{\beta(\mu-\delta)}},
    \qquad
    Q_{00}'=
    \frac{1-\mathrm{e}^{\beta(\mu-\delta)}}
        {1+\mathrm{e}^{\beta\mu}+\mathrm{e}^{\beta(\mu-\delta)}}
    \label{eq3-04}
\end{equation}
in the zero approximation with respect to hopping.

\begin{wrapfigure}{i}{0.5\textwidth}
\centerline{\includegraphics[width=0.47\textwidth]{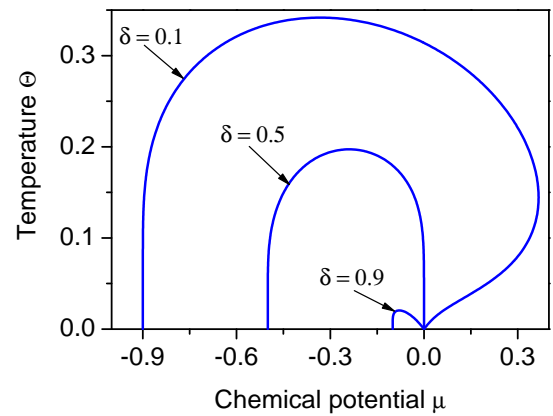}}
\caption{Lines of the NO phase instability (spinodals) with
respect to the appearance of BE condensate in the $(\Theta,\mu)$
plane in the HCB limit at various excitation energies ($t_0=0$,
$\abs{t_0'}=1$, $t_0''=0$).}
\label{fig00}
\end{wrapfigure}

We take into account (according to estimations made in
\cite{wrk15,wrk25}) that boson wave functions in adjacent
potential wells overlap in greater extent in the excited states
compared to the ground ones. Accordingly, we shall put here
$t_0=0$. For a centrosymmetric lattice and in the case of
different parity of wave functions of ground and excited states we
have also $t_0''=0$. Finally, we follow a usual convention of the
BH model for optical lattices taking $t_0'<0$. In this way
equation \eqref{eq3-03} can be reduced to
\begin{equation}
    \frac{\abs{t_0'}}{\delta-\mu}
    \frac{1-\mathrm{e}^{\beta(\mu-\delta)}}
        {1+\mathrm{e}^{\beta\mu}+\mathrm{e}^{\beta(\mu-\delta)}}
    =1
    .
    \label{eq3-05}
\end{equation}
Its solutions determine the stability region boundaries of the
normal (NO) phase. Respective lines of spinodals are numerically
calculated and presented in figure~\ref{fig00} (here and below the
energy quantities are given in units of $\abs{t_0'}$).

As illustrated in figure~\ref{fig00}, at $\delta<\abs{t_0'}$
spinodals surround an asymmetric area in the $(\Theta,\mu)$ plane
which is located between the points $\mu=\delta-\abs{t_0'}$ and
$\mu=0$ of the abscissa axis.
In this region, the NO phase is unstable; this is connected with the
appearance of BE condensate.
At $\delta<\abs{t_0'}/2$ and $\mu>0$
the backward path of spinodal is observed and
a lower temperature of the NO phase instability appears, thus
suggesting a possibility of the SF phase existence in the
intermediate temperature range (so-called ``re-entrant
transition''). However, as will be shown further, in the mentioned
region a real thermodynamic behaviour is even more complicated.
The order of the NO-SF transition can change to the first one and
the SF-phase remains stable up to the zero temperature.

\section{Phase diagrams in MFA}

For a more detailed treatment of the NO-SF transition issue, let us
study the thermodynamics of the considered system in the HCB limit,
thus reducing the problem to a three-state model with the
Hamiltonian
\begin{equation}
    \hat{H} = \sum_{ip} \lambda_p X_i^{pp}
    + \sum_{ij} t_{ij} X_i^{10} X_j^{01}
    + \sum_{ij} t_{ij}' X_i^{20} X_j^{02}
    + \sum_{ij} t_{ij}'' (X_i^{10} X_j^{02}+X_i^{20} X_j^{01}),
    \label{eq4-01}
\end{equation}
where the shorthand notations are used
\begin{equation}
    | 0 \rangle \equiv | 00 \rangle,\;
    | 1 \rangle \equiv | 10 \rangle,\;
    | 2 \rangle \equiv | 01 \rangle;
    \qquad
    \lambda_0 =\lambda_{00},\;
    \lambda_1 =\lambda_{10},\;
    \lambda_2 =\lambda_{01}\,.
    \label{eq4-02}
\end{equation}

Possibility of BE condensation will be studied in the MFA. Average
values of creation (annihilation) operators for Bose particles in
the ground or excited local state
\begin{equation}
    \eta = \langle X_i^{10} \rangle = \langle X_i^{01} \rangle
        \;(\equiv\langle b_i \rangle),
    \qquad
    \xi  = \langle X_i^{20} \rangle = \langle X_i^{02} \rangle
        \;(\equiv\langle c_i \rangle)
    \label{eq4-03}
\end{equation}
play the role of order parameters for the SF-phase. Hence, the
mean-field Hamiltonian is as follows:
\begin{align}
    \hat{H}_{\mathrm{MF}} &=
    -N(t_0 \eta^2 + t_0' \xi^2 + 2 t_0'' \eta\xi)
    + \sum_{ip} \lambda_p X_i^{pp}
    \notag\\
    &\quad
    + \sum_{i}
    \left[
    t_0 \eta (X_i^{10}+X_i^{01})
    + t_0' \xi (X_i^{20}+X_i^{02})
    + t_0'' \xi (X_i^{10}+X_i^{01})
    + t_0'' \eta(X_i^{20}+X_i^{02})
    \right]
    .
    \label{eq4-04}
\end{align}
Self-consistency equations for parameters $\eta$ and $\xi$
\begin{equation}
    \eta = Z^{-1}\Sp[X_i^{10}\exp(-\beta\hat{H}_{\mathrm{MF}})],
    \qquad
    \xi  = Z^{-1}\Sp[X_i^{20}\exp(-\beta\hat{H}_{\mathrm{MF}})]
    \label{eq4-05}
\end{equation}
are equivalent to the condition of minimum of the grand canonical
potential $\Omega=-\Theta \ln Z$, where
$Z=\Sp\exp(-\beta\hat{H}_{\mathrm{MF}})$.

Limiting our consideration to the case of particle hopping only
through excited states ($t_0'\ne0$, $t_0=t_0''=0$) we can
diagonalize Hamiltonian \eqref{eq4-04} by a rotation
transformation
\begin{equation}
    \left(
    \begin{array}{l}
        | 0 \rangle \\
        | 1 \rangle \\
        | 2 \rangle
    \end{array}
    \right)
    =
    \left(
    \begin{array}{ccc}
        \cos\vartheta & 0 & -\sin\vartheta \\
        0 & 1 & 0 \\
        \sin\vartheta & 0 & \cos\vartheta
    \end{array}
    \right)
    \left(
    \begin{array}{l}
        | \tilde{0} \rangle \\
        | \tilde{1} \rangle \\
        | \tilde{2} \rangle
    \end{array}
    \right),
    \label{eq4-06}
\end{equation}
where
\begin{equation}
    \cos2\vartheta =
    \frac{\lambda_2-\lambda_0}{\sqrt{(\lambda_2-\lambda_0)^2+4(t_0'\xi)^2}},
    \qquad
    \sin2\vartheta =
    \frac{2\abs{t_0'}\xi}{\sqrt{(\lambda_2-\lambda_0)^2+4(t_0'\xi)^2}}
    \label{eq4-07}
\end{equation}
and $\lambda_2-\lambda_0=\delta-\mu$. In terms of operators
$\widetilde{X}^{rs}=|\tilde{r}\rangle\langle\tilde{s}|$
\begin{equation}
    \hat{H}_{\mathrm{MF}} =
    N\abs{t_0'}\xi^2+\sum_{ip}\tilde{\lambda}_p\widetilde{X}_i^{pp}.
    \label{eq4-08}
\end{equation}
New energies of single-site states are
\begin{equation}
    \tilde{\lambda}_{0,2} =
    \frac{\delta-\mu}{2}\mp
    \sqrt{\left(\frac{\delta-\mu}{2}\right)^2+(t_0'\xi)^2},
    \qquad
    \tilde{\lambda}_1 = -\mu.
    \label{eq4-09}
\end{equation}
In the new basis
\begin{equation}
    X_i^{02}+X_i^{20} =
    -(\widetilde{X}_i^{22}-\widetilde{X}_i^{00})\sin2\vartheta
    +(\widetilde{X}_i^{20}-\widetilde{X}_i^{02})\cos2\vartheta,
    \label{eq4-10}
\end{equation}
which yields after averaging
\begin{equation}
    \xi =
    \frac{1}{2}
    \langle\widetilde{X}_i^{00}-\widetilde{X}_i^{22}\rangle
    \sin2\vartheta.
    \label{eq4-11}
\end{equation}
Taking into account that
$\langle\widetilde{X}^{pp}\rangle=Z^{-1}\exp(-\beta\tilde{\lambda}_p)$,
$Z=\sum_p\exp(-\beta\tilde{\lambda}_p)$
we come to the equation for
\begin{wrapfigure}[19]{i}{0.5\textwidth}
\centerline{\includegraphics[width=0.47\textwidth]{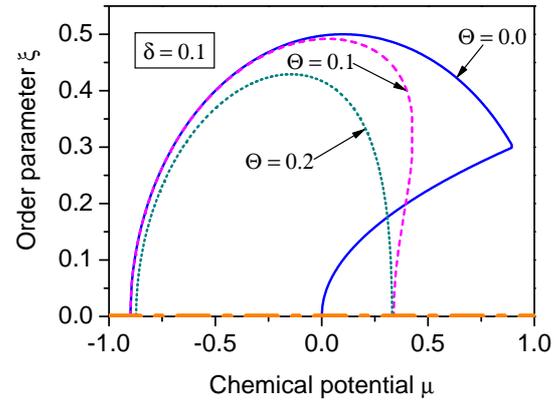}}
\caption{Dependences of the order parameter $\xi$ on the chemical
potential $\mu$ for the reduced three-level (HCB) model at various
temperatures indicating the possibility of the first order phase
transition at low enough temperatures ($\delta=0.1$,
$\abs{t_0'}=1$).}
\label{fig01}
\end{wrapfigure}
the order parameter $\xi$:
\begin{equation}
    \xi =
    \frac{1}{Z}
    \,
    \frac{\abs{t_0'}\xi}{\sqrt{(\delta-\mu)^2+4(t_0'\xi)^2}}
    \left(
    \mathrm{e}^{-\beta\tilde{\lambda}_0}
    -
    \mathrm{e}^{-\beta\tilde{\lambda}_2}
    \right)
    .
    \label{eq4-12}
\end{equation}
Solution $\xi=0$ corresponds to the NO phase. A nonzero solution
describing the BE condensate is obtained from the equation
\begin{equation}
    \frac{1}{Z}
    \,
    \frac{\abs{t_0'}}{\sqrt{(\delta-\mu)^2+4(t_0'\xi)^2}}
    \left(
    \mathrm{e}^{-\beta\tilde{\lambda}_0}
    -
    \mathrm{e}^{-\beta\tilde{\lambda}_2}
    \right)
    = 1
    .
    \label{eq4-13}
\end{equation}
In the limit $\xi\to0$ this equation determines the line where the
order parameter for the SF phase tends to zero. One can readily
see that it coincides with spinodal equation \eqref{eq3-05} thus
defining the line of the second order NO-SF phase transition (when
just the transition of such an order takes place).

\begin{figure}[!b]
\includegraphics[width=0.47\textwidth]{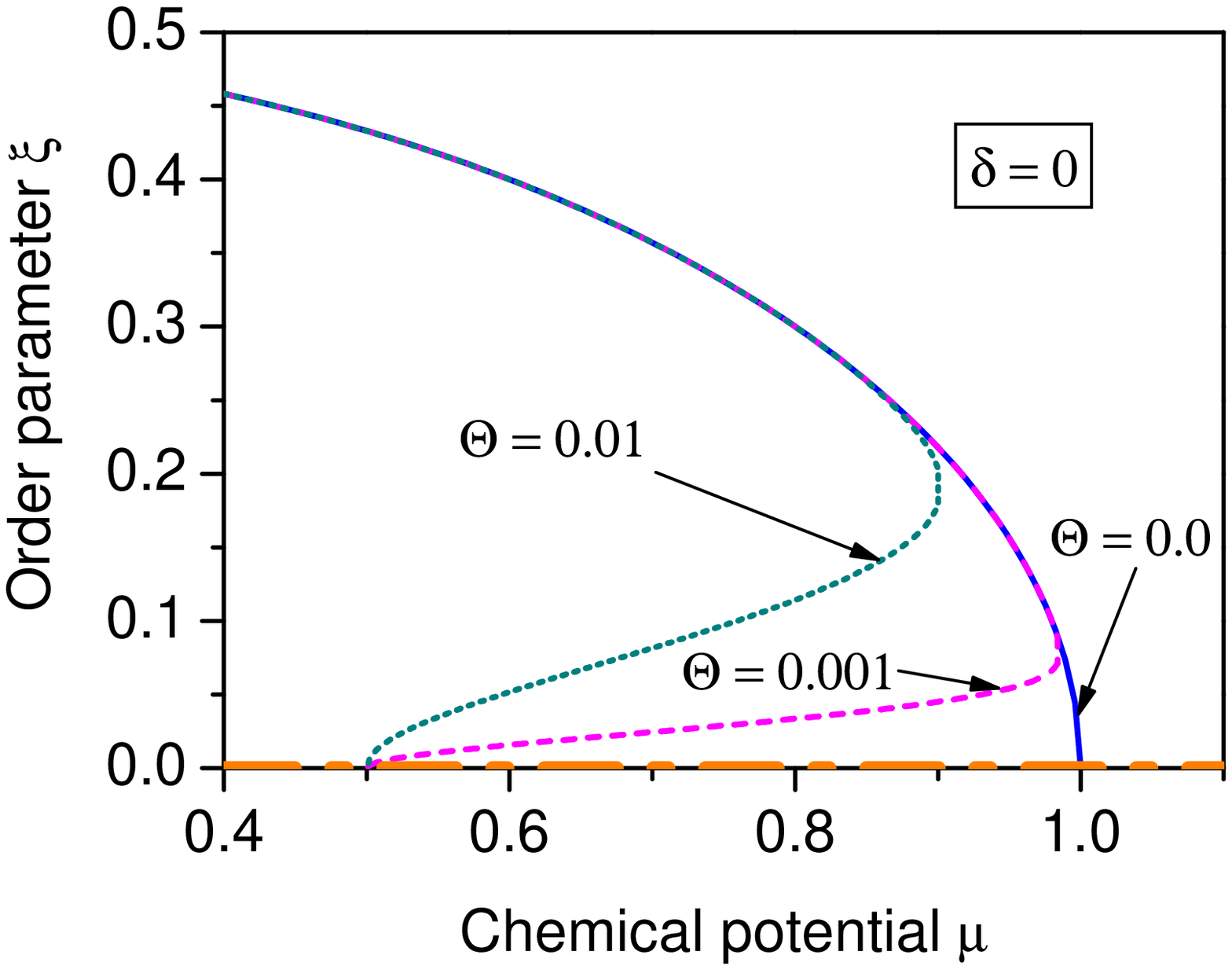}%
\hfill%
\includegraphics[width=0.47\textwidth]{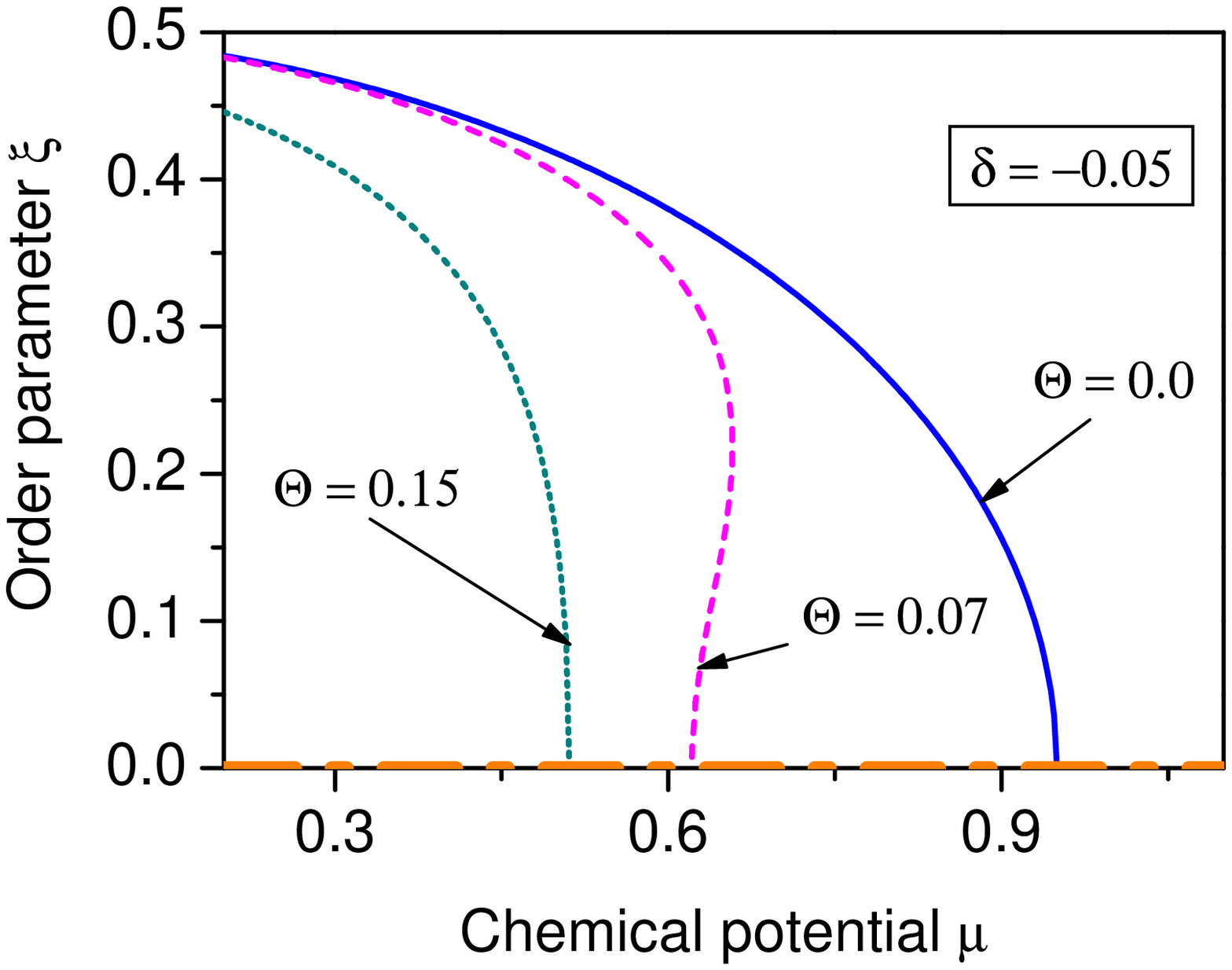}%
\caption{Low-temperature behaviour of the order parameter $\xi$
for the reduced three-level (HCB) model at zero and negative
excitation energies $\delta$ and various temperatures
($\abs{t_0'}=1$).}
\label{fig05}
\end{figure}

Numerical solutions of equation \eqref{eq4-13} make it possible to study the
behavior of the order parameter $\xi$ depending on chemical
potential $\mu$ at various temperatures as illustrated in
figure~\ref{fig01}. In the main, at negative values of chemical
potential the parameter $\xi$ changes smoothly and the phase
transition to the SF phase is of the second order. But at
$\mu\gtrsim0$ and low enough temperatures, the $\xi(\mu)$
dependence has an S-like bend. In this case, the first order phase
transition with an abrupt change of the parameter $\xi$ takes
place. This phase transition occurs at a certain value of the
chemical potential which could be calculated using the Maxwell
rule or considering the minimum of the grand canonical potential
$\Omega(\mu)$ as a function of the chemical potential (see below).
Obviously, the point of $\xi$ nullification does not anymore correspond
here to the phase transition.

Similar behaviour of the parameter $\xi$ holds even at zero
excitation energy ($\delta=0$) where the first order phase
transition remains for nonzero temperatures whereas at $T=0$ its
order changes to the second one (figure~\ref{fig05}). At negative
values of $\delta$ (which corresponds to inversion of
$\varepsilon$ and $\varepsilon'$ levels and to hopping between
ground states) the second order of the transition is preserved in
the low-temperature region close to $T=0$ transforming to the
first order transition at the temperature increase and recovering
henceforth (figure~\ref{fig05}).

\begin{figure}
\includegraphics[width=0.46\textwidth]{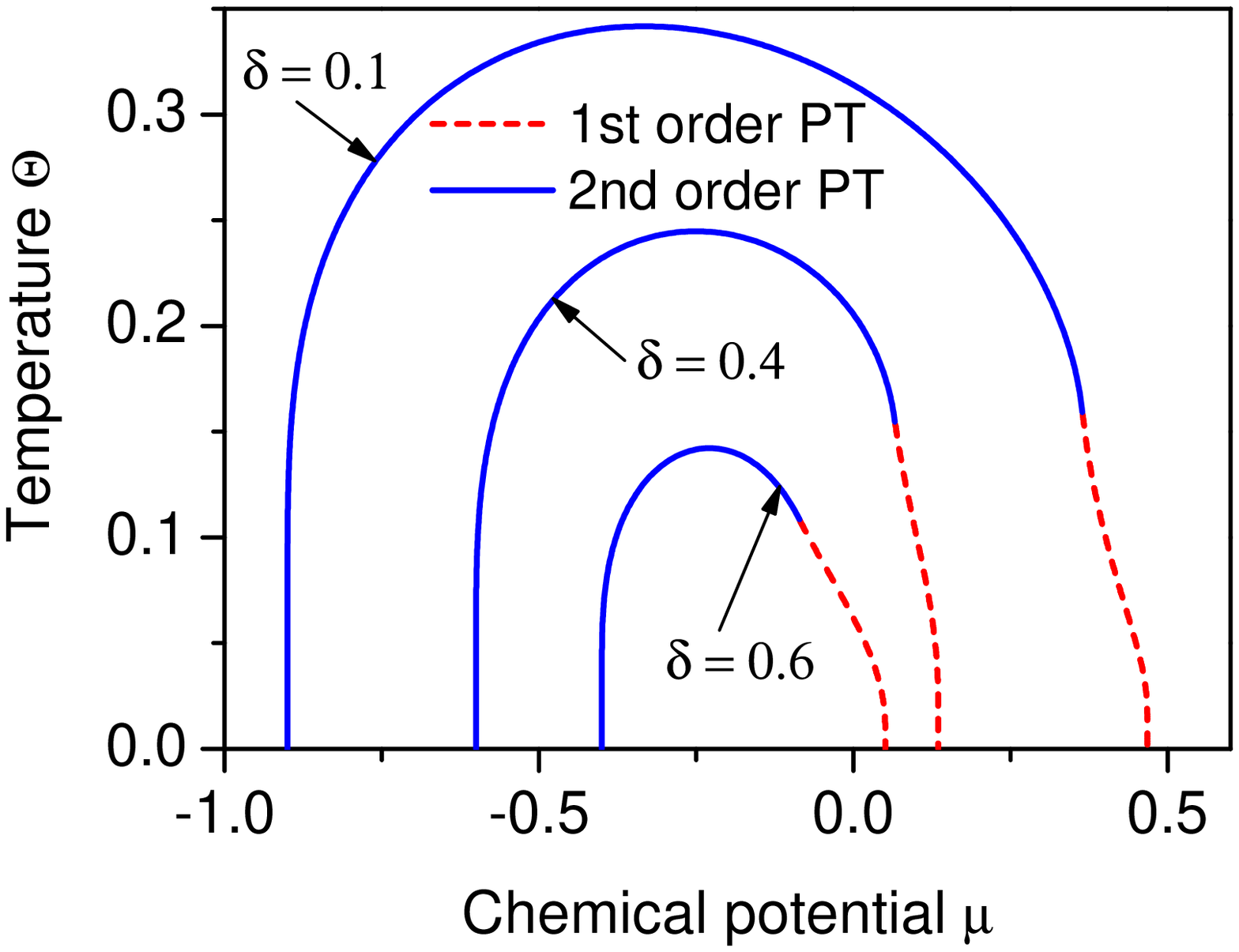}%
\hfill%
\includegraphics[width=0.48\textwidth]{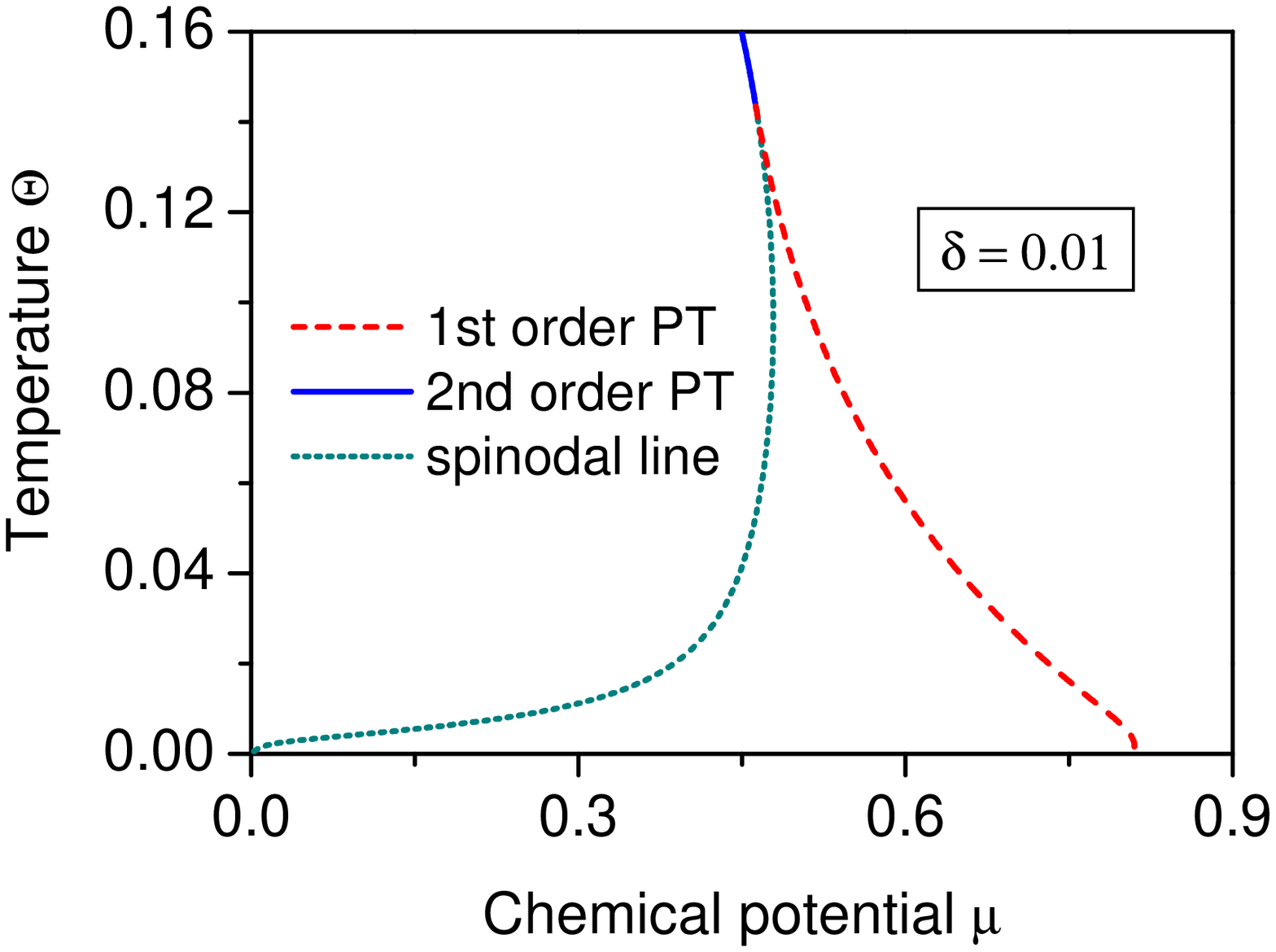}%
\\
\parbox[t]{0.46\textwidth}{%
\caption{Lines of the NO-SF phase transition in the $(\Theta,\mu)$
plane at various excitation
energies $\delta$ ($\abs{t_0'}=1$).}\label{fig06}}%
\hfill%
\parbox[t]{0.5\textwidth}{%
\caption{An illustration of discrepancy between the spinodal curve
and the real line of the first order phase transition for
$\delta=0.01$ ($\abs{t_0'}=1$).}\label{fig07}}%
\end{figure}

\begin{figure}[!b]
\includegraphics[width=0.49\textwidth]{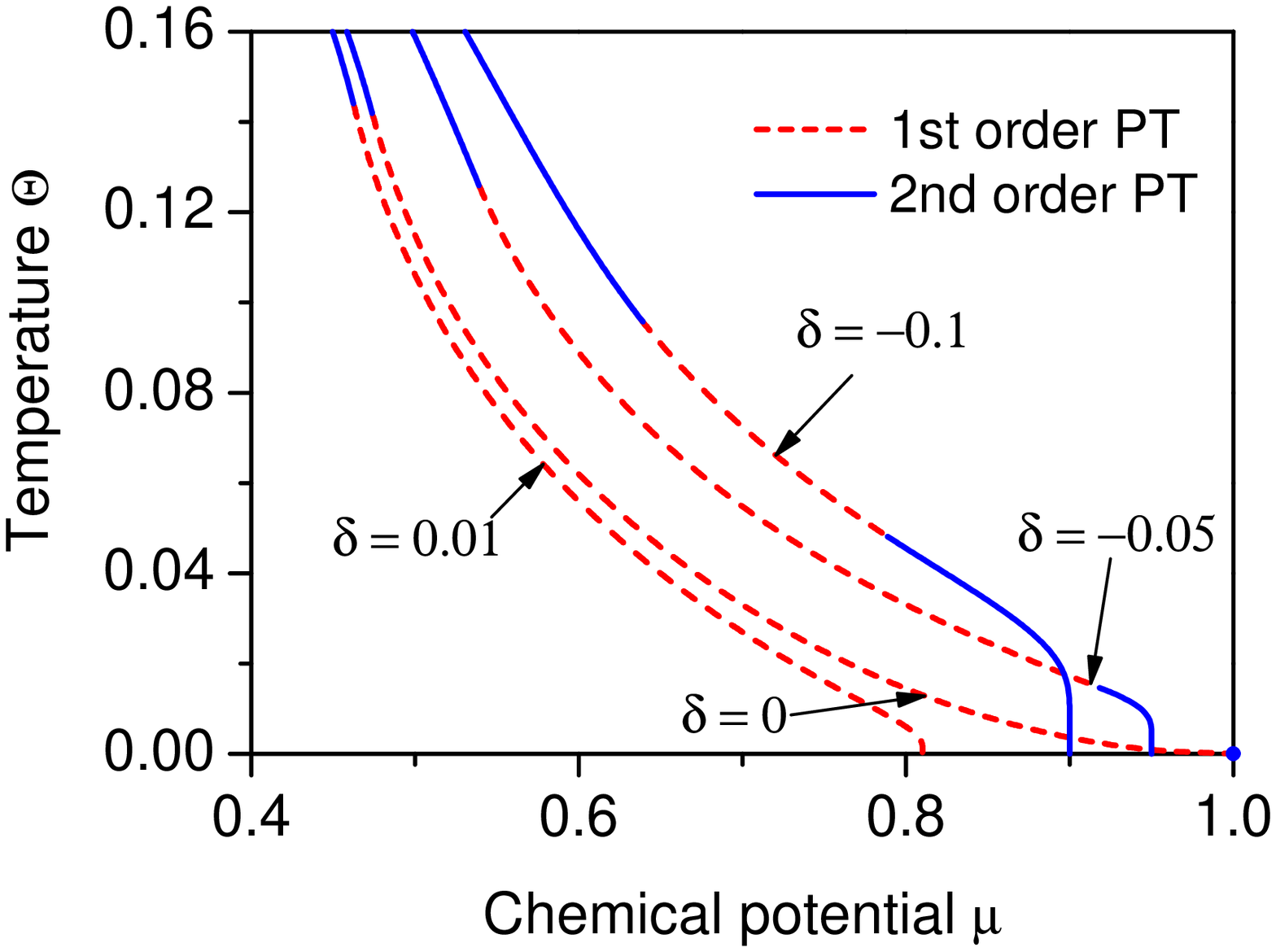}%
\hfill%
\includegraphics[width=0.45\textwidth]{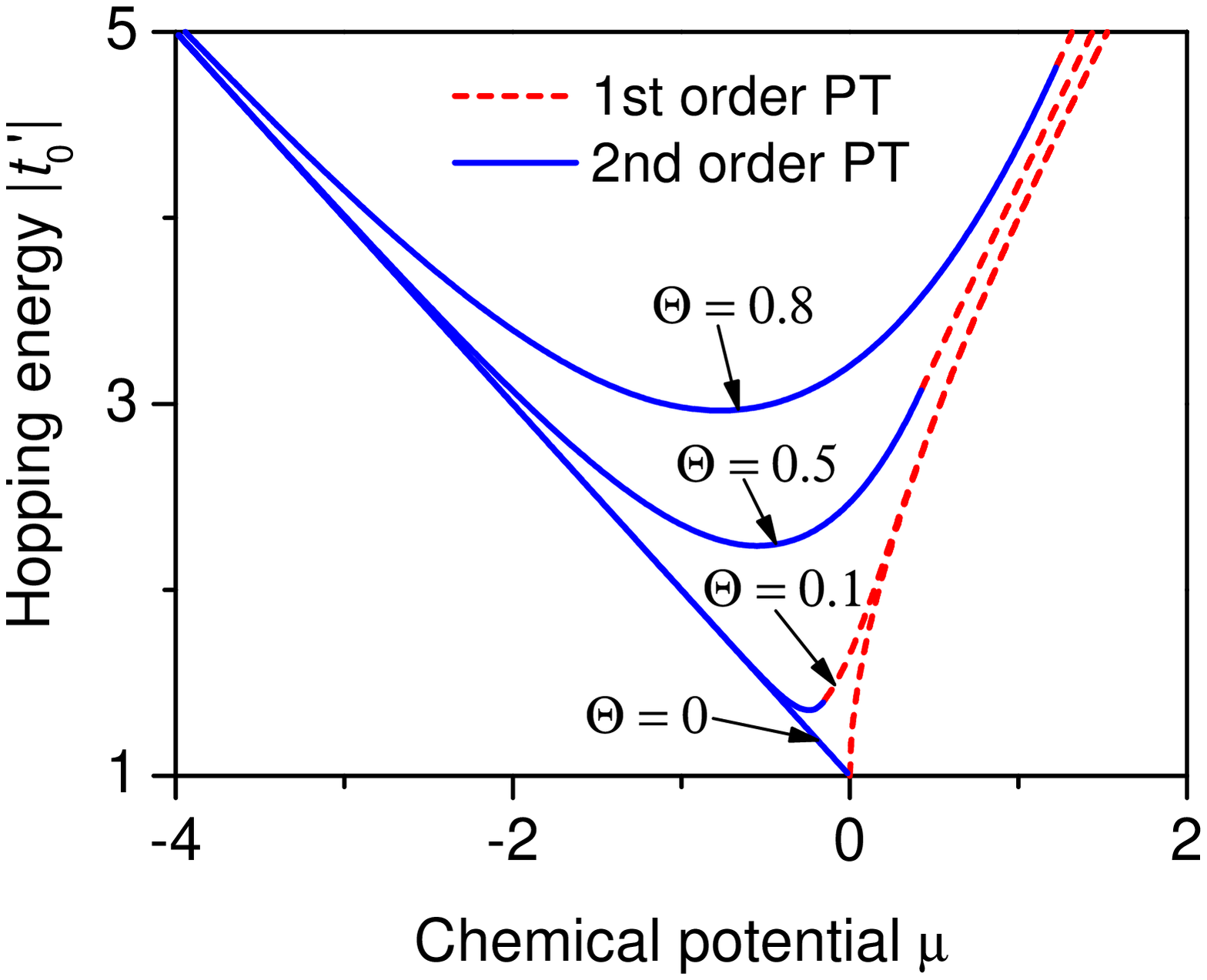}%
\\
\parbox[t]{0.48\textwidth}{%
\caption{Appearance of two tricritical points at zero and negative
values of excitation energy $\delta$ ($\abs{t_0'}=1$).}\label{fig08}}%
\hfill%
\parbox[t]{0.49\textwidth}{%
\caption{Lines of the NO-SF phase transition in the
$(\abs{t_0'},\mu)$ plane at various temperatures $\Theta$ (energy
quantities are given in units of $\delta$).}\label{fig09}}%
\end{figure}

Changes in the NO-SF phase transition order and localization of
the corresponding tricritical points are depicted in
figure~\ref{fig06}, where phase diagrams are given for various
values of the excitation energy $\delta$. At temperatures lower
than tricritical, spinodal lines and phase transition
curves come apart as one can see comparing figures~\ref{fig00}
and~\ref{fig06}. At small values of $\delta$, the discrepancy is
quite significant (figure~\ref{fig07}). In the case of $\delta<0$,
two critical points appear at a certain distance; the latter tends
to zero at $\delta=\delta_{\mathrm{crit}}\approx-0.12\abs{t_0'}$
and the first order phase transitions at a further increase of
$\delta$ (figure~\ref{fig08}) is suppressed.

Phase diagrams in the $(\abs{t_0'},\mu)$ plane at various
temperatures for $\delta>0$ are depicted in figure~\ref{fig09}
with indication of tricritical points. In distinction to the
standard two-level HCB model~\cite{wrk38} (where the SF phase
transition is of the second order) the diagrams are asymmetric. In
the limit $T\to0$ for $\mu>0$ the first order transition occurs at
$\mu=(\sqrt{\delta}-\sqrt{\abs{t_0'}})^2$ (see the next section)
whereas for $\mu<0$ they are of the second order on the line
$\mu=\delta-\abs{t_0'}$.

\section{Phase separation at fixed boson concentration}

Let us consider now the thermodynamics of the model at a fixed
concentration of Bose particles. We will utilize a connection
between the concentration and the chemical potential of bosons
which can be established using its definition in such a form
\begin{align}
    n &\equiv \langle n_i^b + n_i^c \rangle
        = \langle X_i^{11} + X_i^{22} \rangle
    \label{eq5-01}\\
    \intertext{or basing on the relationship}
    n &= -\frac{\partial(\Omega/N)}{\partial\mu}\,.
    \label{eq5-02}
\end{align}

In the first case similarly to equality \eqref{eq4-10} one can
obtain a relation
\begin{equation}
    X_i^{11} + X_i^{22}
    =
    \widetilde{X}_i^{11}
    +\widetilde{X}_i^{00} \sin^2\vartheta
    +\widetilde{X}_i^{22} \cos^2\vartheta
    +(\widetilde{X}_i^{02}+\widetilde{X}_i^{20}) \sin\vartheta \cos\vartheta
    \label{eq5-03}
\end{equation}
which results in
\begin{align}
    n &= \langle \widetilde{X}_i^{11} \rangle
        + \langle \widetilde{X}_i^{00} \rangle \sin^2\vartheta
        + \langle \widetilde{X}_i^{22} \rangle \cos^2\vartheta =
    \notag\\
    &= \frac{1}{Z}
    \left\{
        \mathrm{e}^{-\beta\tilde{\lambda}_1}
        +
        \left[
        \frac{1}{2}-
        \frac{\delta-\mu}{2\sqrt{(\delta-\mu)^2+4(t_0'\xi)^2}}
        \right]
        \mathrm{e}^{-\beta\tilde{\lambda}_0}
        +
        \left[
        \frac{1}{2}+
        \frac{\delta-\mu}{2\sqrt{(\delta-\mu)^2+4(t_0'\xi)^2}}
        \right]
        \mathrm{e}^{-\beta\tilde{\lambda}_2}
    \right\}.
    \label{eq5-04}
\end{align}
In the second case, taking into account that
\begin{equation}
    \Omega / N
    =
    \abs{t_0'}\xi^2 - \Theta \ln Z,
    \qquad
    Z
    =
    \mathrm{e}^{\beta\mu}
    +
    \mathrm{e}^{-\beta(\delta-\mu)/2}
    \cosh\beta\sqrt{\bigl(\textstyle\frac{\delta-\mu}{2}\bigr)^2+(t_0'\xi)^2}
    \label{eq5-05}
\end{equation}
and differentiating with respect to $\mu$, one can come to the
same expression as \eqref{eq5-04}.

There are different relationships between $n$ and $\mu$ in NO and
SF phases; in the last case, a nonzero value of $\xi$ (a solution
of equation \eqref{eq4-13}) should be substituted into expression
\eqref{eq5-04}. Order parameter $\xi$ has a jump at the first
order phase transition, so a stepwise change of concentration $n$
takes place. In the $n=\mathrm{const}$ regime (at the value of $n$
in the region of step) it means a phase separation into two phases
with different concentrations: the NO phase ($\xi=0$ and a larger
concentration of bosons) and the SF phase ($\xi\neq0$ and their
smaller concentration).

\begin{figure}[!t]
\includegraphics[width=0.46\textwidth]{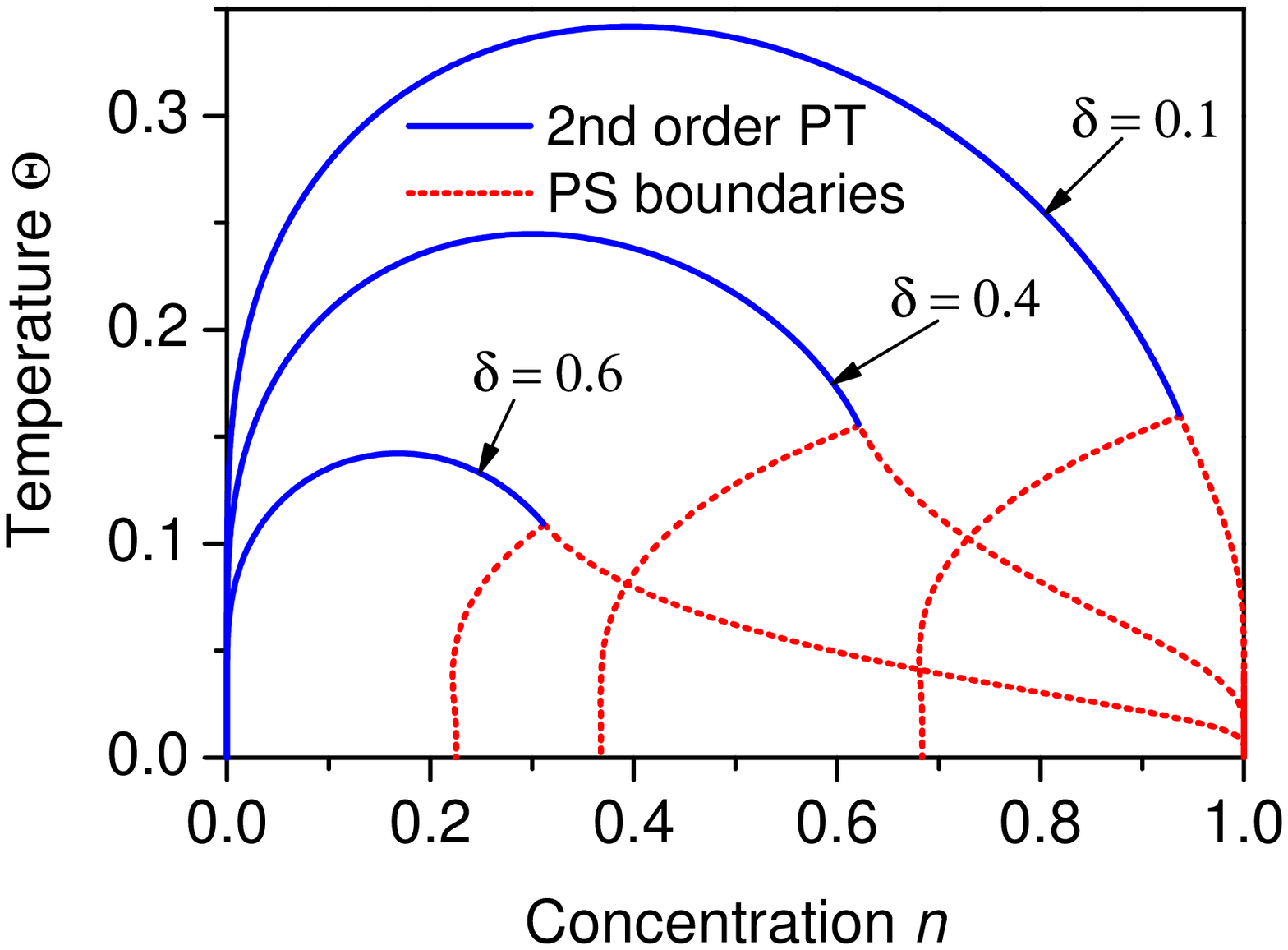}%
\hfill%
\includegraphics[width=0.48\textwidth]{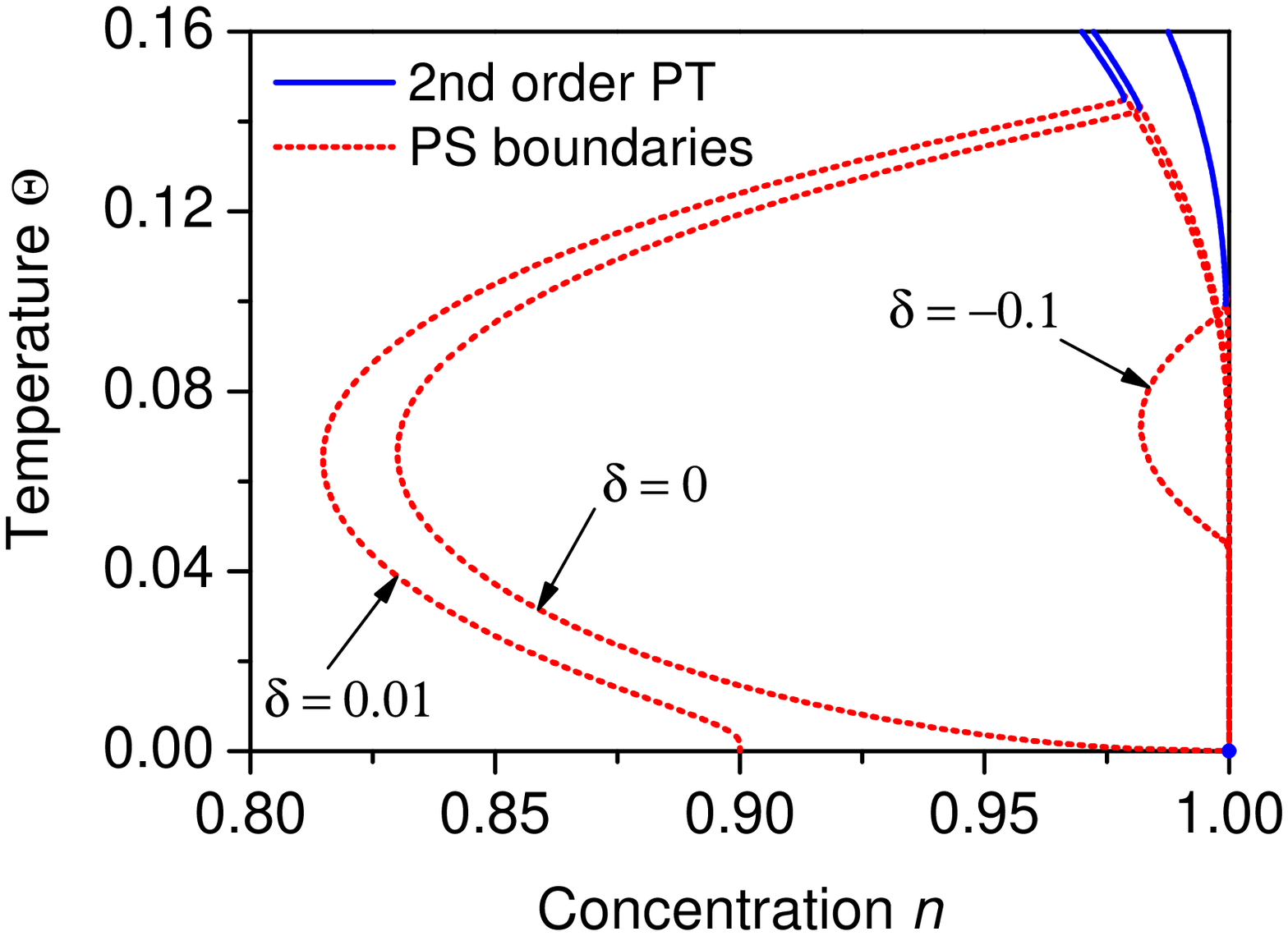}%
\caption{Lines of the NO-SF phase transition and the phase
separation region in the $(\Theta,n)$ plane at various excitation
energies $\delta$ including the case of small, zero and negative
values of $\delta$ ($\abs{t_0'}=1$).}
\label{fig11}
\end{figure}

The above described situation is illustrated in
figure~\ref{fig11}, where the numerically calculated $(\Theta,n)$
phase diagrams are presented. At $\delta>0$, phase separation
region spans up to tricritical temperatures. When $\delta$ goes to
zero and finally reverses its sign, the shape of the separation
region changes in a peculiar way moving off abscissa axis
(figure~\ref{fig11}). Now the phase separation begins at nonzero
temperatures and vanishes at $\delta<\delta_{\mathrm{crit}}$; the
line of the second order phase transition remains only. At the
further increase of $\abs{\delta}$ (in the $\mu<0$ region) the
$(\Theta,n)$ diagram becomes more and more symmetric, approaching
by its shape  the diagram known for the usual HCB model
\cite{wrk39} (see also~\cite{wrk40}).

\begin{figure}[!b]
\includegraphics[width=0.47\textwidth]{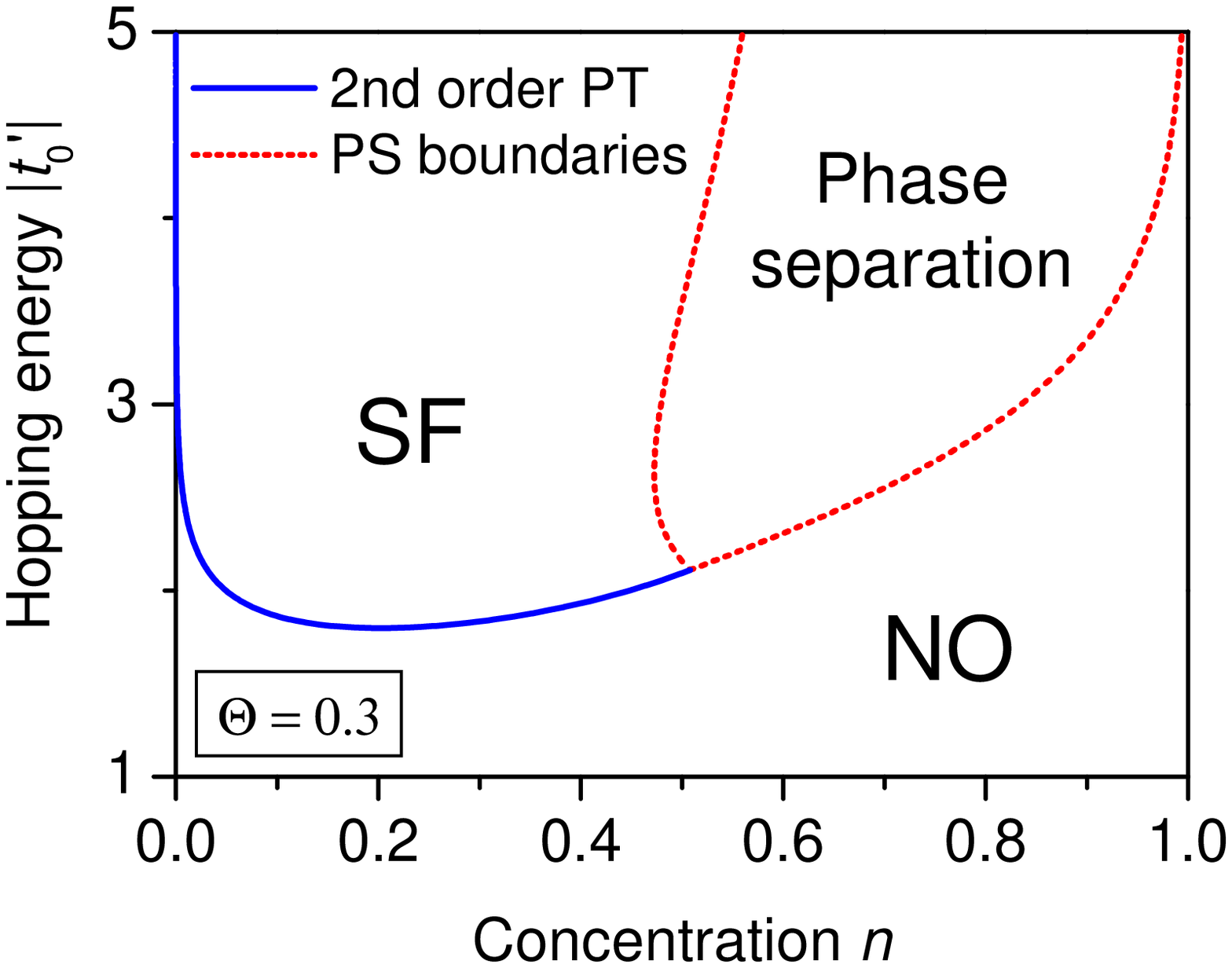}%
\hfill%
\includegraphics[width=0.47\textwidth]{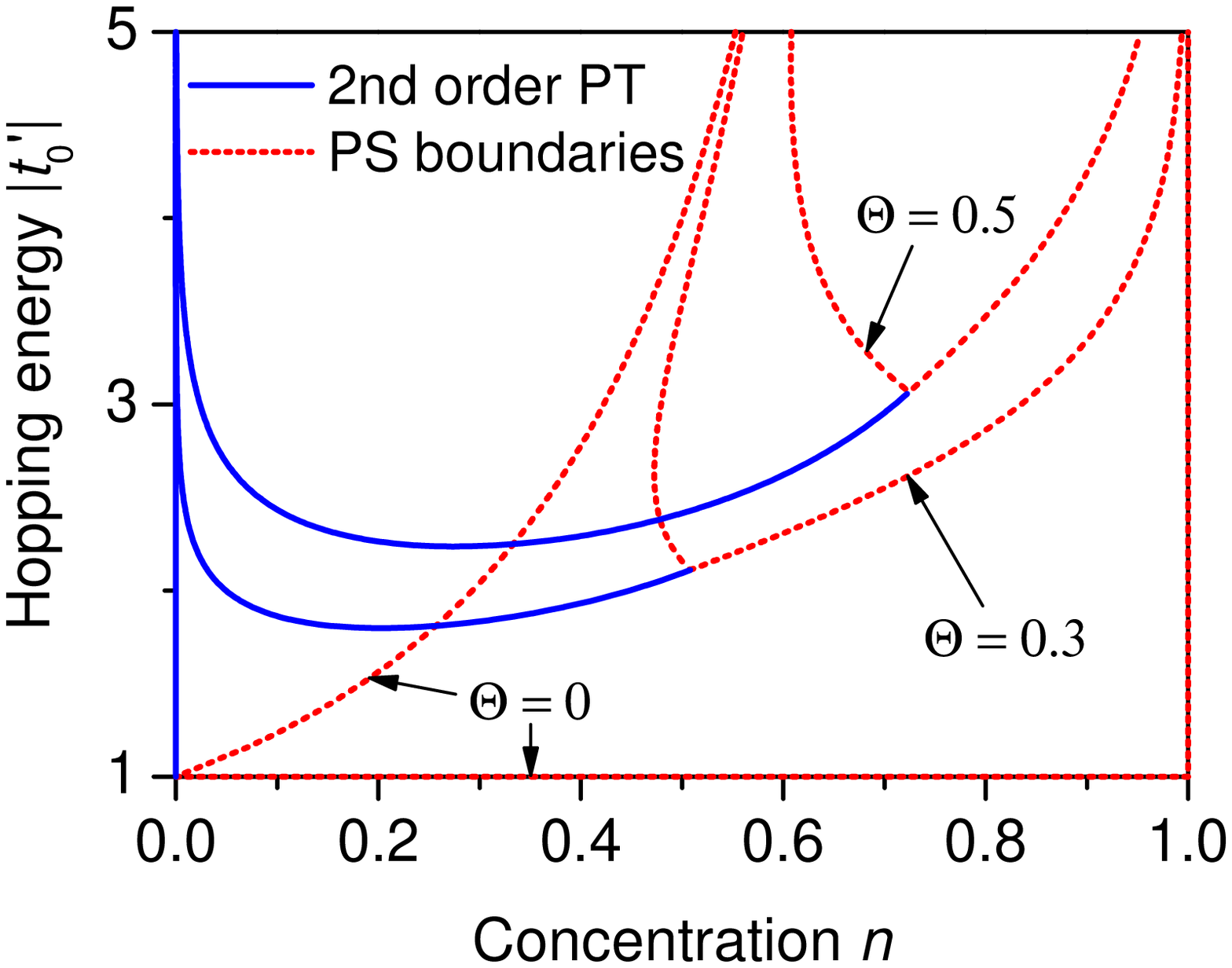}%
\caption{Phase diagram with the indication of possible phases
(above) and lines of the NO-SF phase transition in the
$(\abs{t_0'},\mu)$ plane at various temperatures $\Theta$ (energy
quantities are given in units of $\delta$).}
\label{fig12}
\end{figure}

Phase diagrams in the $(\abs{t_0'},n)$ coordinates are given in
figure~\ref{fig12} where the regions of NO, SF and separated
phases are shown at various temperatures.

The case of the zero temperature can be studied  more in detail in
a pure analytic way. In this limit there are three branches of
order parameter $\xi$ as a function of the chemical potential (see
figure~\ref{fig01}):
\begin{align}
    (1) &\colon \xi=\frac{1}{2\abs{t_0'}}
        \sqrt{\abs{t_0'}^2-(\mu-\delta)^2},
    \notag\\
    (2) &\colon \xi=\sqrt{\mu\delta}/\abs{t_0'},
    \notag\\
    (3) &\colon \xi=0.
    \label{eq5-07}
\end{align}
After elimination of $\xi$ parameter, one can obtain the grand
canonical potential $\Omega$ as follows:
\begin{align}
    (1) &\colon \Omega/N=
        \frac{(\mu-\delta+\abs{t_0'})^2}{4\abs{t_0'}},
    \notag\\
    (2) &\colon \Omega/N= (\delta/\abs{t_0'}-1)\mu,
    \notag\\
    (3) &\colon \Omega/N=
    \begin{cases}
        0,& \mu<0,\\
        -\mu,& \mu>0.
    \end{cases}
    \label{eq5-08}
\end{align}
Differentiating expressions \eqref{eq5-08} with respect to $\mu$
we have
\begin{align}
    (1) &\colon n= \frac{1}{2}+\frac{\mu-\delta}{2\abs{t_0'}},
    \notag\\
    (2) &\colon n= 1-\delta/\abs{t_0'},
    \notag\\
    (3) &\colon n=
    \begin{cases}
        0,& \mu<0,\\
        1,& \mu>0.
    \end{cases}
    \label{eq5-09}
\end{align}

At the first order phase transition from the SF phase to the NO
phase, the order parameter $\xi$ jumps from branch~(1) to branch~(3).
This occurs at the
$\mu=\mu^*\equiv(\sqrt{\abs{t_0'}}-\sqrt{\delta})^2$
value given by equality of respective grand canonical potentials
$\Omega_{(1)}=\Omega_{(3)}$. Then boson system separates into SF
and NO phases with concentrations of bosons:
\begin{align}
    n_{\mathrm{SF}}&= \frac{1}{2}+\frac{\mu^*-\delta}{2\abs{t_0'}},
    \notag\\
    n_{\mathrm{NO}}&= 1.
    \label{eq5-10}
\end{align}

\section{Discussion and conclusions}

As was shown in this work, the transition to the SF phase (the
phase with BE condensate) in the Bose-Hubbard model with two local
states (the ground and excited ones) on the lattice site can be of
the first order in the case, when the particle hopping takes place
only in the excited band. Calculations and estimates for optical
lattices give evidence of significant distinction between hopping
parameters $t_0$ and $t_1$ in the ground and excited bands,
respectively. It follows from estimates~\cite{wrk25} that
$t_1/t_0\approx30-50$ depending on depth $V_0$ of local potential
wells (one can produce effect on $V_0$ changing the intensity of
laser beams which create an optical lattice). Similar results are
obtained in the studies of quantum delocalization of the adsorbed
hydrogen atoms. One can see from calculations~\cite{wrk22,wrk23}
of energy spectrum of the H-atom subsystem on the nickel surface
that the ground-state band has a negligible bandwidth. At the same
time, for excited bands, the bandwidth varies in the range from 15
to 45~meV (depending on the excited state symmetry and on the
crystallographic orientation of metal surface), being mostly of
the order of half the corresponding excitation energy
$\Delta\varepsilon_{\alpha}=\varepsilon_{\alpha}-\varepsilon_0$\,.
There are, however, the cases of strong delocalization (e.g.\ H on
the Ni(110) surface) where the excited bands overlap, and the
width of the lowest one is of the same order as
$\Delta\varepsilon_{\alpha}$~\cite{wrk22}.

The values of hopping parameters greatly increase at the decrease
of $V_0$; the distance between the local energy levels becomes
smaller in this case (see~\cite{wrk33,wrk34}). It is one of the
possible ways of changing the relation between the hopping
parameters and excitation energy ($\abs{t_0'}$ and $\delta$ in our
model). Another possibility (discussed in~\cite{wrk35}) is
connected with an essential reduction of the energy gap between
local $s$- and $p$-levels due to sufficiently strong interspecies
Feshbach resonance in the presence of Fermi atoms added to the
Bose system in optical lattice.

Along with investigation of BE condensation in the excited band
(or bands $p_x,p_y$ ($p_x,p_y,p_z$) in two- (three-) dimensional
case) on condition that certain concentration of Bose-atoms has
been created in the band by optical pumping~\cite{wrk25,wrk34}, an
attempt was made in~\cite{wrk36} to study the effect of excited
bands on the physics of BE condensation in the lowest ($s$-) band
(when the $s$-band hopping is taken into account). The case of
finite values of the one-site interaction $U$ was considered. The
possibility of the re-entrant behaviour of the MI-SF transition
was claimed. However, the order of phase transition was not
investigated; the consideration was restricted to the case of zero
temperature. As we show in this work, re-entrant type dependence
on $T$ or $\mu$ takes place only for spinodals and the return to
the initial MI phase from the SF phase could be possible only in
the case of the second order phase transitions. In reality, the
order of phase transition changes to the first order in this
region. In the HCB limit (no more than one particle per lattice
site), it takes place mainly at positive values of chemical
potential of particles; at $\mu<0$, the transition remains, for
the most part, of the second order. The region of existence of SF
phase is restricted, as a whole, to the interval
$-\abs{t_0'}<\mu<\abs{t_0'}$, while excitation energy should obey
the inequality $\delta<\abs{t_0'}$. We have constructed the
corresponding phase diagrams and established localization of
tricritical points, where the order of phase transition changes.
The separation on SF and NO phases at the fixed particle
concentration is investigated; the conditions of the appearance of
phase-separated state are analyzed.

It should be mentioned that phase diagrams in
figures~\ref{fig01}--\ref{fig12} are close by their shape to the
diagrams obtained in the framework of Bose-Hubbard model for Bose
atoms with spin $S=1$ in optical lattices~\cite{wrk29}. The
excited levels are formed in that case by the higher spin
single-site states and corresponding interactions of the
``ferromagnetic'' or ``antiferromagnetic'' type (the
Hund-rule-like splitting), while the hopping parameter is taken
the same for all bands. The similarity of the mentioned diagrams
points out to the fact that the role of the excited states in the
change of the phase transition order in going to the phase with
the BE condensate is the same in both cases. Distinction, however,
consists in another genesis of the single-site spectrum. In our
model, in the limiting case of HCB there are no effects connected
with the level splitting due to interaction; the excited
single-particle states are taken by us into account instead.

The consideration developed in this work can be extended to the systems
with the close or degenerate excited local levels. Generalization
of the model by adding inter-site interactions is also
important. It could even make it possible to take into consideration other
phases (density-modulated or supersolid) besides NO and SF ones.

We finally emphasize that the hopping parameter $t_{ij}'$ in the
excited band can be positive; in particular, this concerns the
$p$-bands~\cite{wrk35}. In such a situation, the condensation
takes place into states with wave vector $\vec{Q}$ on the boundary
of the Brillouin zone, while the order parameters $\langle c_Q
\rangle$, $\langle c_Q^+ \rangle$ describe the modulated
condensate. Since $t_Q'=-t_0'$\,, the results obtained in this work
are also valid (with $\abs{t_Q'}$ in place of $\abs{t_0'}$) in
that case.



\begin{thebibliography}{99}

\bibitem{wrk01}
Greiner M., Mandel O., Esslinger T., H{\"a}nsch T.W., Bloch I.,
Nature, 2002,
  \textbf{415}, 39; \\
  \bibdoi{10.1038/415039a}.

\bibitem{wrk02}
Greiner M., Mandel O., H{\"a}nsch T.W., Bloch I., Nature, 2002,
\textbf{419},
  51; \bibdoi{10.1038/nature00968}.

\bibitem{wrk03}
Jaksch D., Bruder C., Cirac J.I., Gardiner C.W., Zoller P., Phys.
Rev. Lett.,
  1998, \textbf{81}, 3108; \\
  \bibdoi{10.1103/PhysRevLett.81.3108}.

\bibitem{wrk04}
Sheshadri K., Krishnamurthy H.R., Pandit R., Ramakrishnan T.V.,
Europhys.
  Lett., 1993, \textbf{22}, 257; \\
  \bibdoi{10.1209/0295-5075/22/4/004}.

\bibitem{wrk05}
Konabe S., Nikuni T., Nakamura M., Phys. Rev. A, 2006,
\textbf{73},
  033621; \\
  \bibdoi{10.1103/PhysRevA.73.033621}.

\bibitem{wrk06}
Ohashi Y., Kitaura M., Matsumoto H., Phys. Rev. A, 2006,
\textbf{73},
  033617; \\
  \bibdoi{10.1103/PhysRevA.73.033617}.

\bibitem{wrk07}
Freericks J.K., Monien H., Europhys. Lett., 1994, \textbf{26},
 545; \bibdoi{10.1209/0295-5075/26/7/012}.

\bibitem{wrk08}
Iskin M., Freericks J.K., Phys. Rev. A, 2009, \textbf{79}, 053634;
\bibdoi{10.1103/PhysRevA.79.053634}.

\bibitem{wrk11}
Byczuk K., Vollhardt D., Phys. Rev. B, 2008, \textbf{77}, 235106;
\bibdoi{10.1103/PhysRevB.77.235106}.

\bibitem{wrk12}
Anders P., Gull E., Pollet L., Troyer M., Werner P., Phys. Rev.
Lett., 2010,
  \textbf{105}, 096402; \\
  \bibdoi{10.1103/PhysRevLett.105.096402}.

\bibitem{wrk09}
Batrouni G.G., Scalettar R.T., Phys. Rev. B, 1992, \textbf{46},
 9051; \bibdoi{10.1103/PhysRevB.46.9051}.

\bibitem{wrk10}
Capogrosso-Sansone B., S{\"o}yler {\c{S}}.G., Prokof'ev N.,
Svistunov B., Phys.
  Rev. A, 2008, \textbf{77}, 015602; \\
  \bibdoi{10.1103/PhysRevA.77.015602}.

\bibitem{wrk13}
Astaldi C., Bianco A., Modesti S., Tosatti E., Phys. Rev. Lett.,
1992,
  \textbf{68}, 90; \\
  \bibdoi{10.1103/PhysRevLett.68.90}.

\bibitem{wrk14}
Nishijima M., Okuyama H., Takagi N., Aruga T., Brenig W., Surface
Science
  Reports, 2005, \textbf{57}, 113; \\
  \bibdoi{10.1016/j.surfrep.2005.03.001}.

\bibitem{wrk15}
Reilly P.D., Harris R.A., Whaley K.B., J. Chem. Phys., 1991,
\textbf{95},
  8599; \bibdoi{10.1063/1.461239}.

\bibitem{wrk16}
Ignatyuk V.V., Phys. Rev. E, 2009, \textbf{80}, 041133;
\bibdoi{10.1103/PhysRevE.80.041133}.

\bibitem{wrk17}
Velychko O.V., Stasyuk I.V., Condens. Matter Phys., 2009,
\textbf{12},
  249.

\bibitem{wrk18}
Mysakovych T.S., Krasnov V.O., Stasyuk I.V., Ukr. J. Phys., 2010,
\textbf{55},
  228.

\bibitem{wrk19}
Micnas R., Ranninger J., Robaszkiewicz S., Rev. Mod. Phys., 1990,
\textbf{62},
  113; \\
  \bibdoi{10.1103/RevModPhys.62.113}.

\bibitem{wrk20}
Mahan G.D., Phys. Rev. B, 1976, \textbf{14}, 780;
\bibdoi{10.1103/PhysRevB.14.780}.

\bibitem{wrk21}
Stasyuk I.V., Dulepa I.R., J. Phys. Studies, 2009, \textbf{13},
 2701 (in Ukrainian).

\bibitem{wrk22}
Puska M.J., Nieminen R.M., Surface Science, 1985, \textbf{157},
413; \bibdoi{10.1016/0039-6028(85)90683-1}.

\bibitem{wrk23}
Brenig W., Surface Science, 1993, \textbf{291}, 207;
\bibdoi{10.1016/0039-6028(93)91492-8}.

\bibitem{wrk24}
M{\"u}ller T., F{\"o}lling S., Widera A., Bloch I., Phys. Rev.
Lett., 2007,
  \textbf{99}, 200405; \\
  \bibdoi{10.1103/PhysRevLett.99.200405}.

\bibitem{wrk25}
Isacsson A., Girvin S.M., Phys. Rev. A, 2005, \textbf{72}, 053604;
\bibdoi{10.1103/PhysRevA.72.053604}.

\bibitem{wrk26}
Demler E., Zhou F., Phys. Rev. Lett., 2002, \textbf{88}, 163001;
\bibdoi{10.1103/PhysRevLett.88.163001}.

\bibitem{wrk27}
Krutitsky K.V., Graham R., Phys. Rev. A, 2004, \textbf{70},
063610; \bibdoi{10.1103/PhysRevA.70.063610}.

\bibitem{wrk28}
Kimura T., Tsuchiya S., Kurihara S., Phys. Rev. Lett., 2005,
\textbf{94},
  110403; \\
  \bibdoi{10.1103/PhysRevLett.94.110403}.

\bibitem{wrk29}
Pai R.V., Sheshadri K., Pandit R., Phys. Rev. B, 2008,
\textbf{77},
  014503; \bibdoi{10.1103/PhysRevB.77.014503}.

\bibitem{wrk30}
Chen G.-H., Wu Y.-S., Phys. Rev. A, 2003, \textbf{67}, 013606;
\bibdoi{10.1103/PhysRevA.67.013606}.

\bibitem{wrk31}
Hubbard J., Proc. R. Soc. Lond. A, 1965, \textbf{285}, 542.

\bibitem{wrk32}
Haley S.B., Erd{\"o}s P., Phys. Rev. B, 1972, \textbf{5}, 1106;
\bibdoi{10.1103/PhysRevB.5.1106}.

\bibitem{wrk37}
Ohliger M., Pelster A., Green's function approach to the
{Bose}-{Hubbard}
  model, Preprint arXiv:0810.4399v1 [cond-mat.stat-mech], 2008, 4 p.

\bibitem{wrk38}
Schmid G., Todo S., Troyer M., Dorneich A., Phys. Rev. Lett.,
2002,
  \textbf{88}, 167208; \\
  \bibdoi{10.1103/PhysRevLett.88.167208}.

\bibitem{wrk39}
Pedersen M.H., Schneider T., Phys. Rev. B, 1996, \textbf{53},
 5826; \bibdoi{10.1103/PhysRevB.53.5826}.

\bibitem{wrk40}
Stasyuk I.V., Mysakovych T.S., Condens. Matter Phys., 2009,
\textbf{12},
  539.

\bibitem{wrk33}
Bloch I., Dalibard J., Zwerger W., Rev. Mod. Phys., 2008,
\textbf{80},
  885; \bibdoi{10.1103/RevModPhys.80.885}.

\bibitem{wrk34}
Scarola V.W., Das~Sarma S., Phys. Rev. Lett., 2005, \textbf{95},
 033003; \bibdoi{10.1103/PhysRevLett.95.033003}.

\bibitem{wrk35}
Liu W.V., Wu C., Phys. Rev. A, 2006, \textbf{74}, 013607;
\bibdoi{10.1103/PhysRevA.74.013607}.

\bibitem{wrk36}
Larson J., Collin A., Martikainen J.-P., Phys. Rev. A, 2009,
\textbf{79},
  033603; \\
  \bibdoi{10.1103/PhysRevA.79.033603}.

\end{thebibliography}

%
%

\ukrclosing{stasyuk_done_ukr}
\end{document}